\documentstyle[prd,aps,psfig,axodraw]{revtex}

\begin{document}

\title{\Large{Quark Number Susceptibility in  Hard Thermal Loop 
Approximation}}

\author{Purnendu Chakraborty, Munshi G. Mustafa}

\address{Theory Group, Saha Institute of Nuclear Physics, 1/AF Bidhan Nagar,
Kolkata 700 064, India}

\author{Markus H. Thoma}

\address{Max-Planck-Institut f\"ur extraterrestrische Physik,
Giessenbachstra{\ss}e, 85748 Garching, Germany}

\maketitle

\vspace{0.4in}

\begin{abstract}
We calculate the quark number susceptibility in the deconfined phase
of QCD using the hard thermal loop (HTL) approximation for the quark
propagator and quark-meson vertices. 
This improved perturbation theory takes into account important
medium effects such as thermal quark masses and Landau damping in the
quark-gluon plasma. We explicitly show that the Landau damping part in the
quark propagator for space-like quark momenta does not contribute to the quark
number susceptibility due to the quark number conservation. We find that
the quark number susceptibility only due to the collective quark modes 
deviates from the free one around the critical temperature but 
approaches free results at infinite temperature limit. The results are in 
conformity with recent lattice calculations.

\end{abstract}

\vspace{0.4in}

In recent years substantial experimental
and theoretical efforts have been undertaken to investigate the versatile
physics issues involved in ultra-relativistic heavy-ion collisions, i.e.,
collisions of atomic nuclei in which the center-of-mass energy per nucleon is
much larger than the nucleon rest mass. The principal goal of this initiative
is to explore the phase structure of the underlying theory of strong
interactions -- Quantum Chromodynamics (QCD) -- by creating in the
laboratory a new state of matter, the so-called Quark-Gluon Plasma (QGP).
This new state of matter is predicted to exist
under extreme conditions like at high temperatures and/or
densities, when a phase transition takes place from a hadronic to a
deconfined state of quarks and gluons~\cite{mhw}. 
 Such information has essentially been confirmed by numerical lattice QCD
calculations~\cite{lattice} at finite temperature, which show a rapid increase
in energy density and entropy density as a function of temperature.
Numerical solutions of QCD also suggest that the critical temperature 
is about 170 MeV~\cite{karsch} and provide information on the
equation of state~\cite{karsch1}.

The various measurements taken at CERN SPS within the Lead Beam Programme 
do lead to strong $`$circumstantial evidence' for the formation of the 
QGP~\cite{qm99,qm01}. Evidence
is circumstantial as any direct formation of the QGP cannot be identified. 
Only by some noble indirect diagnostic probes like the suppression of the
$J/\Psi$ particle,
the enhanced production of strange particles, especially strange antibaryons,
excess production of photons and dileptons, the formation of
disoriented chiral condensates, etc. the discovery can be achieved.
An extensive amount of theoretical study has also been devoted over the
last two decades in favour of these well accepted probes
of a  deconfinement (QGP) phase.

Recently screening and fluctuation of conserved quantities have been 
considered as important and relevant probes of the QGP formation 
in heavy-ion collisions~\cite{mcler,roland,asakawa,koch}. In the 
confined/chirally broken phase charges are associated with the hadrons 
in integer units whereas in the deconfined/chirally restored phase they
are associated with the quarks in fractional units which could lead to
charge fluctuations which are different in the two phases~\cite{mcler,koch}. 
The fluctuations can generally be related to the 
associated susceptibilities~\cite{mcler,hatsuda}. The quark number 
susceptibility is associated with the number fluctuation which 
measures the response of the number density
with infinitesimal change of the quark chemical potential. 
Hence the quark number susceptibility
can be
related to charge fluctuations~\cite{koch} and is therefore 
of direct experimental relevance.  The quark number 
susceptibility has been investigated in lattice QCD simulation~\cite{gotlieb}
which showed that it is zero at low temperature and rises suddenly to 
nonzero values across
the deconfinement phase transition. At high temperature QCD it has 
been analysed~\cite{zahed} to show non-perturbative temperature
effects at next-to-leading order. Recently, it has been 
discussed~\cite{prakash} in connection with 
the role of the fluctuations during the dense 
stages of the collision to exploit the electromagnetic probes with the 
hadronic probes. A very recent lattice simulation~\cite{gavai} has verified
a new relation between susceptibilities and screening masses and 
explains that the non-perturbative phenomena are closely connected with
deviations from weak coupling limit or bare perturbation theory, indicating
the need to resum the weak coupling series. The purpose of the present
calculation is to investigate the quark number susceptibility within 
the HTL-resummed perturbation theory which incorporates non-perturbative
effects such as effective masses of the collective quark modes 
(quark and plasmino modes in the medium originating from the
poles of the HTL-propagator) and Landau
damping for space-like quark momenta, reflecting the physical picture of
the QGP as a gas of quasiparticles. As we will see below, the quark number 
susceptibility obtained in HTL approximation is in agreement with recent 
lattice~\cite{gavai} observations.

\subsection{ Fluctuation and Susceptibility:}

Let \( {\cal O}_{\alpha } \) be a Heisenberg operator. In a static and uniform
external field \( {\cal F}_{\alpha } \), the (induced) 
expectation value of the operator \( {\cal O}_\alpha \left( 0,\overrightarrow{x}
\right) \) is written~\cite{hatsuda} as 
\begin{equation}
\phi _{\alpha }\equiv \left\langle {\cal O} _{\alpha }\left
( 0,\overrightarrow{x}\right) \right\rangle _{F}=\frac{{\rm Tr}\left
[ {\cal O} _{\alpha }\left( 0,\overrightarrow{x}\right) e^{-\beta \left
( {\cal H}+{\cal H}_{ex}\right) }\right] }{{\rm Tr}\left[ e^{-\beta 
\left( {\cal H}+{\cal H}_{ex}\right) }
\right] }=\frac{1}{V}\int d^{3}x\, \left\langle {\cal O} _{\alpha }
\left( 0,\overrightarrow{x}\right) \right\rangle \: , \label{eq1}
\end{equation}
where translational invariance is assumed and 
\({\cal H}_{ex} \) is given by
\begin{equation}
{\cal H}_{ex}=-\sum _{\alpha }\int d^{3}x\, {\cal O} _{\alpha }\left( 0,
\overrightarrow{x}\right) {\cal F}_{\alpha }\: .\label{eq2}
\end{equation}

The (static) susceptibility \( \chi _{\alpha \beta } \) is defined as

\begin{eqnarray}
\chi _{\alpha \sigma }(T) & = & \left. \frac{\partial \phi _{\alpha }}
{\partial {\cal F}_{\sigma }}\right| _{{\cal F}=0}\nonumber \\
 & = & \beta \int d^{3}x\, \left\langle {\cal O} _{\alpha }\left
( 0,\overrightarrow{x}\right) {\cal O} _{\sigma }\left( 0,\overrightarrow{0}
\right) \right\rangle \: , \label{eq3}
\end{eqnarray}
assuming no broken symmetry \( \left\langle {\cal O} _{\alpha }
\left( 0,\overrightarrow{x}\right) \right\rangle =\left\langle 
{\cal O} _{\sigma }
\left( 0,\overrightarrow{0}\right) \right\rangle =0 \). 
$\langle {\cal O}_\alpha (0,{\vec x}){\cal O}_\sigma(0,{\vec 0})\rangle $
is the two point correlation function with operators evaluated
at equal times.

\subsection{{Quark Number Susceptibility:}}

The quark number susceptibility is the measure of the response of the 
quark number density with infinitesimal changes in the quark chemical 
potential $\mu_q +\delta \mu_q$.
Under such a situation the external field, ${\cal F}_\alpha$, in 
({\ref{eq2}) 
can be identified as the change in quark chemical potential $\mu_q$ and 
the operator ${\cal O}_\alpha$ as 
$j_0 = \overline{q} \gamma _{0}q$, where $j_\mu(t,{\vec x})=
\overline{q} \gamma _{ \mu}q$ is the vector meson current. 
Then the quark number susceptibility for a given quark flavour follows from
(\ref{eq3}) as
\begin{eqnarray}
\chi_q(T) &=& \left.\frac{\partial \rho_q}{\partial \mu_q}\right |_{\mu_q=0} 
\nonumber \\
&=& \beta \int \ d^3x \ \left \langle j_0(0,{\vec x})j_0(0,{\vec 0})
\right \rangle \ =\beta\int \ d^3x \ S_{00}(0,{\vec x}), 
\label{eq4}
\end{eqnarray} 
where $S_{00}(0,{\vec x})$ is the time-time component of the vector 
meson correlator 
$S_{\mu\nu}(t,{\vec x})= \langle j_\mu(t,{\vec x})j_\nu(0,{\vec 0})\rangle$
and  the number density can be written  as 
\begin{equation}
\rho_q=\frac{1}{V} 
\frac{{\rm{Tr}}\left [  {\cal N}_q e^{-\beta \left({\cal H}-\mu_q {\cal N}_q
\right )}\right ]}
{{\rm{Tr}}\left [e^{-\beta \left({\cal H}-\mu_q {\cal N}_q\right )}\right ]} =
\frac{\langle {\cal N}_q\rangle}{V} = -\frac{1}{V} \frac{\partial \Omega}
{\partial \mu_q} \ , \label{eq5}
\end{equation}
with the quark number operator, ${\cal N}_q=\int j_0(t, {\vec x}) \ d^3x$, and 
$\Omega=-T \ln Z$ is the thermodynamic potential and $Z$ the partition
function of a quark-antiquark gas.

Taking the Fourier transform of $S_{00}(0,{\vec x})$, it can be shown that
\cite{hatsuda}
\begin{equation}
\chi _{q}\left( T\right) =\lim_{p\rightarrow 0}\>
\beta \int _{-\infty }^{+\infty }\frac{d\omega }{2\pi }
S_{00}\left(\omega ,p\right) \, \label{eq6}.
\end{equation}
Using the fluctuation-dissipation theorem~\cite{fluc}, it can further 
be shown that
\cite{hatsuda}
\begin{equation}
\chi _{q}\left( T\right) =\beta \int _{-\infty }^{+\infty }\frac{d\omega }
{2\pi }\, \frac{-2}{1-e^{-\beta \omega }}\, \rm{Im}\, 
\Pi _{00}\left( \omega ,0\right) \, , \label{eq7}
\end{equation}
where $\Pi_{\mu\nu}(\omega, {\vec p}) ={\rm {FT}} \left (-i\theta(t) \left \langle
\left [j_\mu(t,{\vec x}),j_\nu(0,{\vec 0})\right ]_{-}\right \rangle
\right )$ and FT stands for Fourier transformation.

\subsubsection{Free case}

\vspace{0.2in}
\begin{center}
\begin{picture}(300,100)(0,0)
\ArrowArc(150,50)(40,180,0)
\ArrowArcn(150,50)(40,180,0)
\DashArrowLine(50,50)(110,50){5}
\DashArrowLine(190,50)(250,50){5}
\Text(75,40)[]{$P$}
\Text(225,40)[]{$P$}
\Text(150,105)[]{$K$}
\Text(150,-5)[]{$Q=P-K$}
\end{picture}
\end{center}
\vspace{0.3in}
\hspace{1.8in} {Fig. 1 The self-energy diagram for free quarks.}

\vspace{0.3in}

To lowest order in perturbation theory one has to evaluate the time-time 
component of the 
self energy diagram shown in Fig 1, where the internal quark lines represent
a bare quark propagator $S_f(L)$ which can be expressed in the helicity 
representation for massless case as ($L=(l_0, {\vec l}\, )$, 
$l=|{\vec l}\, |$) \cite{bpy}
\begin{equation}
\label{2.1}
S_{f}\left( l_{0},l\right) =\frac{\gamma ^{0}-\widehat{l}.
\overrightarrow {\gamma }}{2d_{+}\left( L\right) }+\frac{\gamma ^{0}
+\widehat{l}.\overrightarrow{\gamma }}{2d_{-}\left( L\right) }\, 
\end{equation}
with
\begin{equation}
 d_{\pm }\left( l_{0},l\right) =-l_{0}\pm l \ . \label{2.1a} 
\end{equation}
The corresponding spectral function is given by
\begin{equation}
 \rho^f _{\pm }\left( l_0,l\right) =\delta \left( l_0\mp l\right ). 
\label{2.1b} 
\end{equation}

Now the time-time component of the vector meson self-energy in Fig 1
can be written as
\begin{equation}
\label{2.1c}
\Pi ^{00}\left( P\right) =N_{f}N_{c}T\sum _{k_{0}}\int \frac{d^{3}k}{(2\pi )
^{3}}{\rm{Tr}}\left[ S_{f}\left( K\right) \gamma ^{0}S_{f}\left( Q\right) \gamma ^{0}
\right] \, ,
\end{equation}
where, \( Q=P-K \), and $N_f$ and $N_c$ are, respectively,
the number of quark flavours and colours.

Substituting the propagator (\ref{2.1}) in (\ref{2.1c}), 
performing the trace, 
and following the summation relation of Ref.~\cite{bpy}, we extract the
imaginary part as
\begin{eqnarray}
{\rm{Im}}\, \Pi ^{00}\left( \omega,{\vec p}=0\right) & = & 4N_{f}N_{c}\pi \left
( 1-e^{\beta \omega}
\right) \int \frac{d^{3}k}{(2\pi )^{3}}\int dx \int dx^{\prime }\,
 \delta \left( \omega -x -x^{\prime }\right) \nonumber\\
& & n_{F}\left( x \right) n_{F}\left( x^{\prime} \right)
 \rho^f _{+}\left( x ,k\right) \rho^f _{-}\left( x ^{\prime }
,k\right) \, , \label{2.2}
\end{eqnarray}
where $x$ and $x^\prime$ are the energies of the internal quarks and 
$n_F$ is the Fermi distribution function. 
Using (\ref{2.1b}) in (\ref{2.2}), we
get
\begin{equation}
\label{2.3}
{\rm{Im}}\Pi ^{00}\left( \omega ,{ \vec p}=0\right) =4N_{c}N_{f}\pi
\left (1 - e^{\beta \omega}\right ) \ \delta \left( \omega \right)
 \int \frac{d^{3}k}{\left( 2\pi \right) ^{3}}n_{F}\left( k\right)
\left( 1-n_{F}\left( k\right) \right) \,.
\end{equation}
Inserting (\ref{2.3}) in (\ref{eq7}), 
we obtain the quark number susceptibility 
in lowest order perturbation theory~\cite{mcler,hatsuda} 
\begin{equation}
\label{2.4}
\chi^f_q \left( T\right) =4N_{f}N_{c}\beta \int \frac{d^{3}k}{\left( 2\pi 
\right)^{3}}\frac{e^{\beta k}}{\left( 1+e^{\beta k}\right) ^{2}} \, \, .
\end{equation}
Alternatively, considering the lowest order thermodynamic potential
of a quark-antiquark gas~\cite{kapus,lebell}
\begin{equation}
\Omega = - 2N_fN_c T \int \frac{d^3k}{(2\pi)^3}\left [ \beta E_k +
\ln \left (1 + e^{-\beta(E_k-\mu_q)}   \right )
+\ln \left (1 + e^{-\beta(E_k+\mu_q)}   \right ) \right ] \ , \label{2.5}
\end{equation}
one could also arrive at the
same expression using (\ref{eq4}) and (\ref{eq5}). 

Due to the quark number conservation  
${\rm{Im}} \ \Pi(\omega,0)$, as obtained in (\ref{2.3}), 
is proportional to $\delta(\omega)$. This leads to a general relation between
the quark number susceptibility and the time-time component of the polarisation
tensor in the vector channel 
\begin{equation}
 \chi^f_q \left( T\right) =-4N_{f}N_{c}\int \frac{d^{3}k}{\left( 2\pi 
\right)^{3}} \frac{dn_F}{dk} \equiv -{\rm{Re}} \ \Pi_{00}(0,0) \equiv
 (\mu_D^f)^2 \, \, , \label{2.6}
\end{equation}
which provides a connection~\cite{hatsuda,prakash,kapus,lebell,blaiz} 
to the electric screening mass, $\mu_D^f$. This relation is only valid
in lowest order in perturbation theory. 
 
\subsubsection {HTL case}

Now we turn to the estimate of the quark number susceptibility beyond the free 
quark approximation by invoking the in-medium properties of quarks in a
QGP. In the weak coupling limit ($g\ll 1$), a consistent method is to use the  
HTL-resummed quark propagators and HTL quark-meson vertex if the quark 
momentum is soft ($\sim gT$). Using this improved perturbation theory
at least to some extent non-perturbative features of the QGP such as  
effective quark masses and Landau damping are incorporated through the 
effective quantities like quark propagators and the quark-meson vertex. 

Let us, however, note that we do not aim at a complete leading order 
perturbative calculation. Rather we want to study the influence of medium
effects incorporated in the HTL resummed quark propagator. Hence, we will use 
this propagator for the entire momentum range instead of consistently 
distinguishing between soft and hard momenta \cite{BY}. Anyway, since
this distinction is only possible in the weak coupling limit, $g\ll 1$,
it cannot be used in our case, in which we want to compare our results
to QCD lattice calculations. The approach, considered here, 
is in the same spirit as the one for calculating meson correlators in 
the QGP \cite{karsch2}. It should be 
noticed that our results are gauge independent due to the gauge invariance of
the HTL quark propagator.

The HTL-resummed quark propagator, $S^\star(L)$, can be obtained~\cite{bpy}
 from (\ref{2.1}) by replacing $d_\pm(L)$ as 
\begin{equation}
D_{\pm }\left( l_{0},l\right) =-l_{0} {\pm} l+\frac{m^{2}_{q}}
{l}\left[ Q_{0}\left( \frac{l_{0}}{l}\right) \mp Q_{1}\left( \frac{l_{0}}{l}
\right) \right] \, , \label{2.7}
\end{equation}
where the thermal quark mass is given by
$m_q=g(T) T/{\sqrt 6}$, and $Q_n(y)$ is the Legendre 
function of second kind. The HTL-vertex can be obtained~\cite{taylor}
as
\begin{equation}
\label{2.8}
\Gamma ^{\mu }\left( P_{1},P_{2}\right) =\gamma ^{\mu }+m^{2}_{q}
G^{\mu \nu }\gamma _{\nu }
\end{equation}
with
\begin{equation}
\label{2.9}
G^{\mu \nu }\left( P_{1},P_{2}\right) =\int \frac{d\Omega }{4\pi }
\frac{R^{\mu }R^{\nu }}{(R\cdot P_{1})(R\cdot P_{2})}=G^{\mu \nu }
\left( -P_{1},-P_{2}\right) \, ,
\end{equation}
where $R\equiv (-1, {\vec r})$ is a light-like four vector, $R^2=0$. 
The effective propagator and vertex are related via the Ward identity. 

The HTL-spectral function reads \cite{bpy}
\begin{equation}
\rho_\pm(l_0,l) = \frac{l_0^2-l^2}{2m_q^2} \left [ \delta(l_0-\omega_\pm)
+\delta(l_0+\omega_\mp) \right ]
 +\beta_\pm(l_0,l) \Theta(l^2-l_0^2) \label{2.10}
\end{equation}
with 
\begin{equation}
\beta_\pm(l_0,l) = -\frac{m_q^2}{2} \frac{\pm l_0 -l}
{\left [  l (-l_0 \pm l) +m_q^2\left (\pm 1- \frac{\pm l_0 -l}{2l} \ln 
\frac{l+l_0}{l-l_0} \right )\right ]^2 + \left[ \frac{\pi}{2} m_q^2 
\frac{\pm l_0 -l}{l} \right]^2} \, \, \, .  \label{2.11}
\end{equation}
Here the zeros $\omega_\pm(l)$ of $D_\pm(L)$ describe the two branches of the
dispersion
relation of collective quark modes in a thermal medium~\cite{bpy}. 
Furthermore the 
HTL-resummed quark propagator acquires a cut contribution below the
light cone ($l_0^2 < l^2$) as the quark self-energy has a non-vanishing 
imaginary part, which can be related to Landau damping for space-like quark
momenta resulting from interactions of valence quarks with gluons in the
thermal medium. In addition, an explicit temperature dependence only enters 
through $m_q(T)$ as well as through the  strong 
coupling constant, $g^2(T)=4\pi \alpha_s(T)$
with
 \begin{equation}
 \alpha_s(T) = \frac{12\pi}{(33-2N_f)\ln\left (Q^2/\Lambda_0^2\right)} ,
\label{coupling}
\end{equation}
where $\Lambda_0=200$ - 300 MeV.
For the momentum scale $Q$ we take the energy of the lowest Matsubara mode $Q=2\pi T$
\cite{Andersen99}. For checking the sensitivity of the susceptibility
to uncertainties in the coupling constant we will use also $Q=4\pi T$.
It should also be noted that the 
HTL-resummed propagator is chirally symmetric in spite of the appearance 
of an effective quark mass~\cite{bpy}.
 
Now we need to calculate the imaginary part of the 
time-time component of the self energy diagrams 
given in Fig.~2, in which blobs represent the effective quantities. 
The tadpole 
diagram in Fig.~2b is essential to satisfy the transversality condition, 
$P_\mu\Pi^{\mu\nu} (P)=0$. As  
will be seen below  this has a very important effect on the quark number 
susceptibility by partially compensating the cut contribution in diagram
Fig.~2a.

\vspace{0.2in}
\begin{center}
\begin{picture}(300,100)(0,0)
\ArrowArc(150,50)(40,180,270)
\ArrowArcn(150,50)(40,180,90)
\CArc(150,50)(40,0,90)
\CArc(150,50)(40,270,0)
\DashArrowLine(50,50)(110,50){5}
\DashArrowLine(190,50)(250,50){5}
\Vertex(110,50){4}
\Vertex(190,50){4}
\Vertex(150,90){4}
\Vertex(150,10){4}
\Text(75,40)[]{$P$}
\Text(225,40)[]{$P$}
\Text(150,105)[]{$K$}
\Text(150,-5)[]{$Q=P-K$}
\Text(25,50)[]{(a)}
\end{picture}
\end{center}

\vspace{0.3in}

 \begin{center}
\begin{picture}(300,100)(0,0)
\ArrowArcn(150,50)(40,270,90)
\ArrowArcn(150,50)(40,90,270)
\DashArrowLine(50,10)(150,10){5}
\DashArrowLine(150,10)(230,10){5}
\Vertex(150,10){4}
\Vertex(150,90){4}
\Text(75,20)[]{$P$}
\Text(225,20)[]{$P$}
\Text(150,103)[]{$K$}
\Text(25,10)[]{(b)}
\end{picture}
\end{center}
\vspace{0.3in}

\centerline{Fig. 2 The self-energy (a) and tadpole (b) diagrams for
quarks in the HTL-approximation.}

\vspace{0.3in}
Now the time-time component of the self-energy in diagram  Fig.~2a 
can be written as 
\begin{eqnarray}
\Pi^{00}_1\left( P\right)  & = & N_{f}N_{c}T\sum _{k_{0}}\int \frac{d^{3}k}
{(2\pi )^{3}}{\rm{Tr}} \left [S^\star\left( K\right) \Gamma^{0}
\left( K-P,-K;P\right) S^\star \left( Q\right) 
  \Gamma ^{0}\left( P-K,K;-P\right)\right ] \, .\label{2.12}
\end{eqnarray}
The time component of the HTL-vertex can be obtained 
by carrying out the angular integration in (\ref{2.8})
for \( \overrightarrow{p}=0 \) 
\begin{equation}
\label{2.13}
\Gamma ^{0}\left( K-P,K;P\right) =\left( 1-\frac{m^{2}_{q}}{p_{0}k}\delta
Q_{0}\right) \gamma ^{0}+\frac{m_{q}^{2}}{p_{0}k}\delta Q_{1}\widehat{k}.
\overrightarrow{\gamma } \, ,
\end{equation}
where
 \begin{equation}
\label{2.14}
\delta Q_{n}=Q_{n}\left( \frac{k_{0}}{k}\right) -Q_{n}\left( \frac{k_{0}-p_{0}}{k}\right) \, .
\end{equation}
Alternatively, $\Gamma^0(P_1,P_2)$ can also be obtained from the Ward identity
$P_\mu \Gamma^\mu(P_1,P_2;P) = {S^\star}^{-1}(P_1)-{S^\star}^{-1}(P_2)$.

Now performing the traces in (\ref{2.12}), we get
 \begin{equation}
\label{2.15}
\Pi ^{00}_1\left( p_{0},{ \vec p}=0\right) =2N_{f}N_{c}T\sum _{k_{0}}\int 
\frac{d^{3}k}{(2\pi )^{3}}\left[ \frac{\left( a+b\right) ^{2}}{D_{+}
\left( K\right) D_{-}(Q)}+\frac{(a-b)^{2}}{D_{-}\left( K\right) D_{+}
\left( Q\right) }\right] \, ,
\end{equation}
 where 
\begin{eqnarray}
\label{2.16}
 a&=&\left( 1-\frac{m^{2}_{q}}{p_{0}k}\delta Q_{0}\right)\ , \ \ \ \ \ \ \ \   
 b=\frac{m_{q}^{2}}{p_{0}k}\delta Q_{1} \, \, , \nonumber \\  
a\pm b&=&1-\frac{m_{q}^{2}}{p_{0}k}\left\{ Q_{0}\left( \frac{k_{0}}{k}
\right) \left( 1\mp \frac{k_{0}}{k}\right) +Q_{0}\left( \frac{q_{0}}{k}
\right) \left( 1\pm \frac{q_{0}}{k}\right) \right\} \, ,
\end{eqnarray}
 where  \( q_{0}=p_{0}-k_{0} \) and \( Q_{n}\left( -y\right) =\left( -1\right)
 ^{n+1}Q_{n}\left( y\right)  \)
has been used in (\ref{2.16}).
Following the summation formula of Ref.~\cite{bpy} the imaginary part of 
(\ref{2.15}) then can be written as
 \begin{eqnarray}
{\rm{Im}}\,\Pi ^{00}_1\left( \omega,{\vec p}=0\right)  & = & N_{f}N_{c}\pi 
\left( 1-e^{\beta \omega}\right) \int \frac{d^{3}k}{(2\pi )^{3}}\int dx \int 
dx^{\prime }\delta \left(\omega -x -x^{\prime }\right) n_{F}
\left( x \right) n_{F}\left( x^{\prime }\right) \nonumber \\
 &  & \left[  4 \left( 1-\frac{x+x^\prime}{\omega}\right) ^{2} 
\rho _{+}\left(x ,k\right) \rho _{-}\left( x^{\prime }, k\right) 
\right.  \nonumber \\
&-& \left. 4\frac{m^{2}_{q}}{\omega^{2}k}\Theta \left( k^{2}-x^{2}\right) 
  \left\{ \frac{1}{2}\left( 1-\frac{x }{k}\right) \rho _{+}
\left( x^{\prime },k\right) +\frac{1}{2}\left( 1+\frac{x }{k}\right)
\rho _{-}\left(x ^{\prime },k\right) \right\} \right] \, .\label{2.17}
\end{eqnarray}
The contribution of the time-time component of the tadpole diagram 
in Fig. 2b can be written as
\begin{eqnarray}
\Pi^{00}_2\left( P\right)  & = & N_{f}N_{c}T\sum _{k_{0}}\int \frac{d^{3}k}
{(2\pi )^{3}}{\rm{Tr}} \left [S^\star\left( K\right) \Gamma^{00}
\left(-K,K;-P, P\right)\right ] \, ,\label{2.18}
\end{eqnarray}
where the effective HTL four-point function 
can be obtained from the relation
\begin{equation}
P_\mu\Gamma^{\mu\nu}(-K,K;-P,P) = \Gamma^\nu (K-P, -K;P)-\Gamma^\nu(-K-P,K;P). 
\label{2.18a}
\end{equation} 

At \( \overrightarrow{p}=0 \) the four point function  is obtained as
\begin{eqnarray}
\Gamma ^{00}\left( -K,K;-P,P\right)  & = & -\frac{m^{2}_{q}}{p^{2}_{0}k}
\left( \delta Q_{0}+\delta Q^{\prime }_{0}\right) \gamma ^{0}+\frac{m^{2}_{q}}
{p^{2}_{0}k}\left( \delta Q_{1}+\delta Q^{\prime }_{1}\right) \widehat{k}
\cdot \vec{\gamma } \, \, , \label{2.19}
\end{eqnarray}
with 
\begin{eqnarray}
\delta Q^{\prime }_{n} & = & Q_{n}\left( \frac{k_{0}}{k}\right) 
-Q_{n}\left( \frac{k_{0}+p_{0}}{k}\right) . \label{2.20}
\end{eqnarray}
Proceeding exactly the same way
as before, we get the imaginary part of the tadpole
\begin{eqnarray}
{\rm Im}\, \Pi _{{\rm 2}}^{00}(\omega,{\vec p}=0) & = & N_{f}N_{c}\pi 
\left( 1-e^{\beta \omega} \right) \int \frac{d^{3}k}{\left( 2\pi \right)^{3}}
\int dx \int dx^{\prime }\delta \left(\omega-x -x^{\prime }\right) n_{F}
\left(x \right) n_{F}\left(x^{\prime }\right) \\
 &  & 4 \frac{m_q^{2}}{\omega^{2}k}
\Theta \left( k^{2}-x^{2}\right) 
\left[ \frac{1}{2} \left( 1-\frac{x}{k}\right) \rho _{+}\left(x^\prime ,k
\right) +\frac{1}{2} \left( 1+\frac{x}{k}\right) \rho _{-}\left( 
x^\prime ,k\right) \right] \, \, . \label{2.21}
\end{eqnarray}
It can be seen that the tadpole contribution compensates the 
second term of (\ref{2.17}) and the total contribution becomes
 \begin{eqnarray}
{\rm{Im}}\,\Pi ^{00}\left( \omega,{\vec p}=0\right)  & = & 4 N_{f}N_{c}\pi 
\left( 1-e^{\beta \omega}\right) \int \frac{d^{3}k}{(2\pi )^{3}}\int dx \int 
dx^{\prime }\delta \left(\omega -x -x^{\prime }\right) n_{F}
\left( x \right) n_{F}\left( x^{\prime }\right) \nonumber \\
 &  &  \left( 1-\frac{x+x^\prime}{\omega}\right) ^{2} 
\rho _{+}\left(x ,k\right) \rho _{-}\left( x^{\prime }, k\right) \, \, \, . 
  \label{2.22}
 \end{eqnarray}
Now combining (\ref{2.10}), (\ref{2.22}) and (\ref{eq7}), the quark 
number susceptibility in HTL-approximation is obtained as
\begin{eqnarray}
\chi_q^h(T) &=& 4 N_f N_c \beta \int \frac{d^3k}{(2\pi)^3} \left [
\frac{\left (\omega_{+}^2 (k) -k^2 \right )^2}{4m_q^4}n_F(\omega_{+})
\left ( 1-n_F(\omega_{+}) \right ) \right. \nonumber \\
&+& \left. \frac{\left (\omega_{-}^2 (k) -k^2 \right )^2}{4m_q^4}n_F(\omega_{-})
\left ( 1-n_F(\omega_{-}) \right )
\right ] \, \,  . \label{2.23}
\end{eqnarray}
Since the quark number susceptibility is constructed from
two quark propagators, in general they should receive pole-pole, pole-cut
and cut-cut contributions \cite{bpy}. However, according to
(\ref{2.23}) it has only 
pole-pole contributions from the two collective quark modes 
(quark and plasmino modes in the medium). The cut contributions 
(pole-cut and cut-cut) due to space-like quark momenta (Landau damping) 
do not contribute because of the number conservation in the system. In the
high temperature limit, the second term due to the plasmino mode decouples
from the medium whereas the first term reduces to the free susceptibility
given in (\ref{2.6}).

\vspace{0.3in}

 \centerline{\psfig{figure=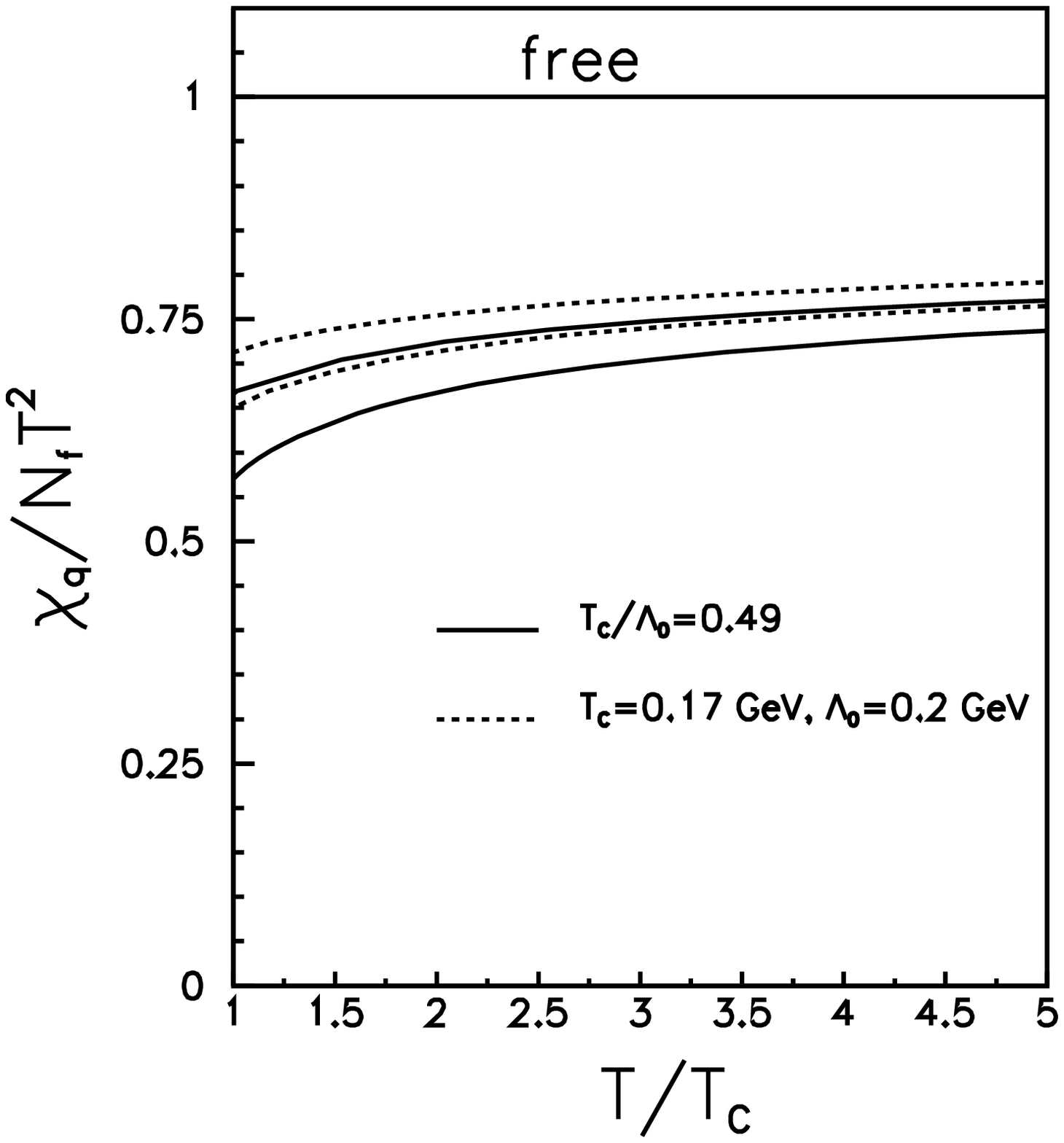,height=10cm,width=10cm}}

\vspace{-0.5in} 

\noindent {Fig 3. $\chi_q(T)/N_f T^2$ as a function of $T/T_c$ in the 
HTL-approximation. $N_f$ is the
number of quark flavours and $T_c$ is the critical temperature for the
deconfined phase transition. For the solid curves 
(\ref{coupling}) with $T_c/\Lambda_0=0.49$ 
\cite{Gupta01} and for the dashed curves $\Lambda_0=200$ MeV and 
$T_c=170$ MeV have been used.
The lower curves correspond to $Q=2 \pi T$ and the upper ones to $Q=4 \pi T$.}

\vspace{0.3in}

We now discuss our results for the quark number susceptibility. In Fig. 3 
we present the temperature dependence of the quark number susceptibility
$\chi_q(T)$ in units of $N_f T^2$, the free field theory susceptibility.
We have used the temperature dependence of the strong coupling constant
as given in (\ref{coupling}) for two different ratios of $T_c$ to $\Lambda_0$
and two different 
values for $Q$. The choice 
$T_c/\Lambda_0=0.49$
\cite{Gupta01} has been used in the lattice calculations of Ref.~\cite{gavai}. 
We observe that the quark number susceptibility does not depend strongly
on the choice of the coupling constant. (For $T_c=170$ MeV and
$\Lambda_0=300$ MeV the susceptibility decreases by about 5\% compared to
the dashed curves.) 
The susceptibility increases with the increase of temperature.  
It is interesting 
to note that the result lies significantly below the free result even at
$T=5T_c$ and the deviation is $25-40\%$. This deviation agrees with
the recent lattice observation~\cite{gavai} along with a  slow
approach to the free quark susceptibility. This result is reminiscent of 
the fact that the quark number susceptibility in HTL-approximation contains 
non-perturbative information about the QGP phase at high temperature.
In the infinite temperature limit, the HTL result exactly reproduces the free 
quark susceptibility. This can clearly be seen in Fig. 4 below. 

In Fig. 4 we display the quark number susceptibility in units of $N_fT^2$
as function of $m_q/T$ using (\ref{coupling}) with 
$\Lambda_0=300$ MeV and
$Q=2\pi T$. As expected for small thermal quark masses,
the HTL-susceptibility begins with 
the free field theory result, $ N_fT^2$, with a flat tangent at 
$m_q=0$ and a quadratic dependence on $m_q$. 
With the increase of the mass the susceptibility starts decreasing
monotonically.  For low values of the thermal quark mass, 
the susceptibility is large due to the fact that it is relatively easy to
create an additional quark or antiquark. 
On the other hand, if quarks acquire thermal masses in the medium, the 
susceptibility decreases, which could qualitatively be 
understood as an effect of the Boltzmann factor. 

 \centerline{\psfig{figure=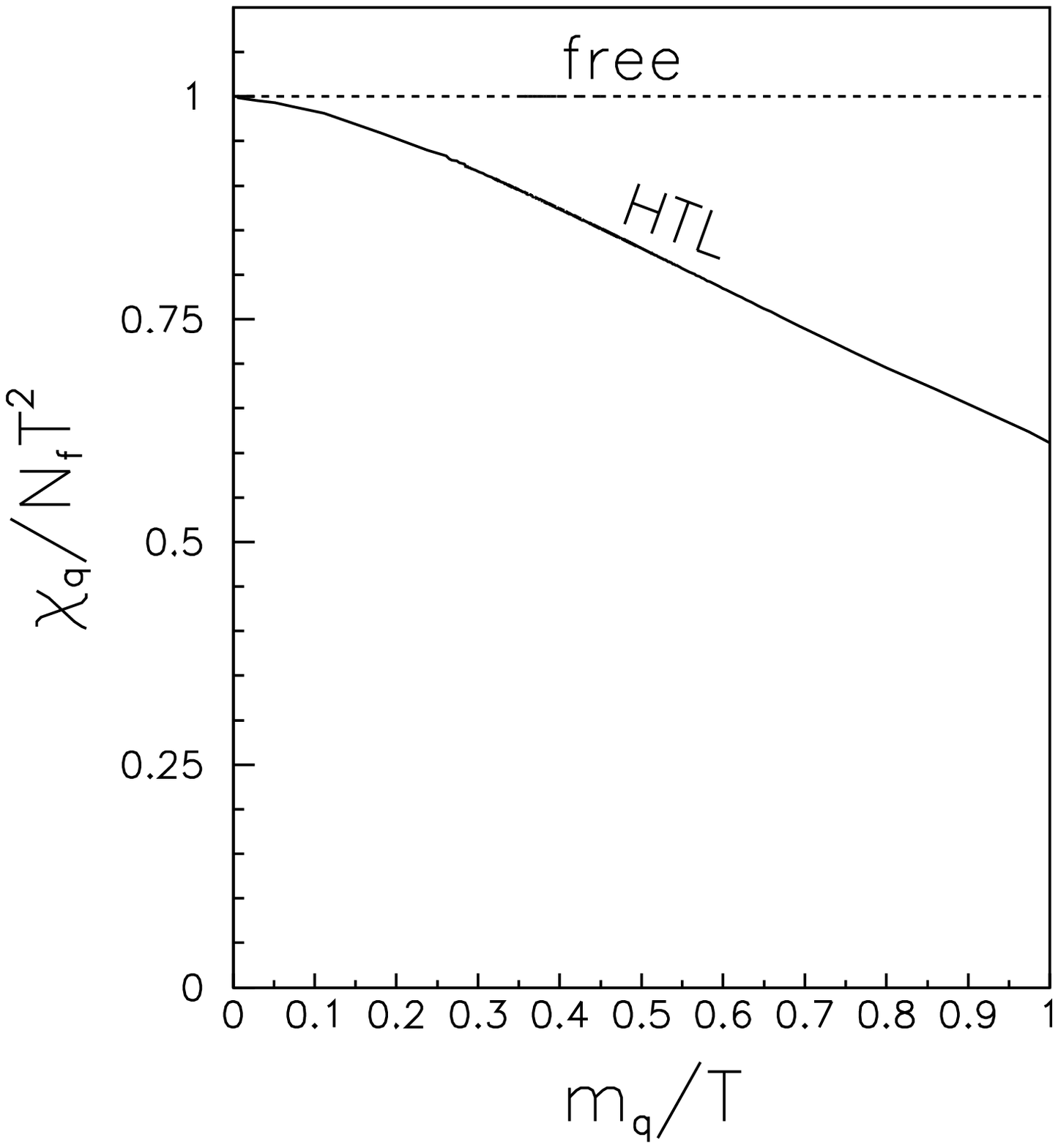,height=10cm,width=10cm}}

\vspace{-0.5in} 

\noindent {Fig 4. $\chi_q(T)/N_f T^2$ as a function of $m_q/T$ for
free quarks (dashed line) and in HTL-approximation (solid line). $m_q$ is 
the thermal quark mass.}

\vspace{0.3in}
\subsection{Conclusion:}

We have calculated the quark number susceptibility in the HTL-approximation
which incorporates in-medium effects of the QGP phase such as quark 
masses and Landau damping through the HTL resummed propagators and HTL 
quark-meson vertex in the vector meson channel. 
We have discussed the influence of the various non-perturbative effects on
the quark number susceptibility in the deconfined phase. The Landau damping
contribution due to space-like  quark momenta drops out because of the
quark number conservation in the system. Technically this occurs partly
due to a cancellation of the two diagrams in Fig.2 containing
HTL-resummed quark-meson vertices, and partly due to kinematical reasons in
(\ref{2.22}). 

We find that the quark number
susceptibility is significantly smaller than 
the free susceptibility for moderately high 
temperatures. Our results are in good agreement with recent 
lattice calculations. 
Besides the on-set of
confinement and chiral symmetry breaking close to $T_c$, HTL effects
may explain the lattice results similar as in the
case of the free energy \cite{bir} but in contrast to
the meson correlation functions \cite{karsch2}.
Since the quark number 
susceptibility is related to charge fluctuations, it could be an
interesting observable in heavy-ion collisions.

\bigskip

{\it Note added:} After completing our investigation a preprint by
Blaizot, Iancu, and Rebhan appeared on the same topic \cite{bir2}.
They computed the quark number susceptibility from the thermodynamic 
potential, obtained in an approximately self-consistent resummation of
HTLs. However, they do not use the effective HTL resummed vertices, which
plays an important role by partly compensating the Landau damping contribution
in the quark number susceptibility as shown in our investigation.
As in the present paper, they found results similar to lattice data.

\bigskip

{\bf Ackowledgements:}

\medskip

We would like to thank S. Gupta and A. Rebhan for helpful comments.
P.C. is thankful to R. Pisarski for useful communication.

\end{document}